\documentclass[aps,twocolumn,showpacs]{revtex4}
\usepackage{epsfig}
\usepackage{graphicx}
\usepackage{textcomp}
\usepackage{epstopdf}

\usepackage{soul}

\usepackage{amsmath, amsthm, amssymb}
\usepackage{bm}

\usepackage[colorlinks=true, urlcolor=blue, anchorcolor=blue, citecolor=blue,filecolor=blue,linkcolor=blue,menucolor=blue,pagecolor=blue]{hyperref}

\newcommand{\vect}[1]{\bm {#1}} 

\usepackage{colordvi}
\usepackage[usenames,dvipsnames]{xcolor}

\begin{document}
\title{Local average height distribution of fluctuating interfaces}

\author{Naftali R. Smith}
\email{naftali.smith@mail.huji.ac.il}
\affiliation{Racah Institute of Physics, Hebrew
  University of Jerusalem, Jerusalem 91904, Israel}

\author{Baruch Meerson}
\email{meerson@mail.huji.ac.il}
\affiliation{Racah Institute of Physics, Hebrew
  University of Jerusalem, Jerusalem 91904, Israel}

\author{Pavel V. Sasorov}
\email{pavel.sasorov@gmail.com}
\affiliation{Keldysh Institute of Applied Mathematics, Moscow 125047, Russia}


\begin{abstract}
Height fluctuations of growing surfaces can be characterized by the probability distribution of height in a spatial point at a finite time.
Recently there has been spectacular progress in the studies of this quantity for the Kardar-Parisi-Zhang (KPZ) equation in $1+1$ dimensions.  Here we notice that, at or above a critical dimension, the finite-time one-point height distribution is ill-defined in a broad class of linear surface growth models, unless the model is regularized at small scales. The regularization via a system-dependent small-scale cutoff leads to a partial loss of universality. As a possible alternative,
we introduce a \emph{local average height}. For the linear models the
probability density of this quantity is well-defined in any dimension. The weak-noise theory (WNT) for these models yields the ``optimal path" of the interface conditioned on a non-equilibrium fluctuation of the local average height. As an illustration, we consider the
conserved Edwards-Wilkinson (EW) equation,
where, without regularization, the finite-time one-point height distribution is ill-defined in all physical dimensions. We also determine the optimal path of the interface in a closely related problem of the finite-time \emph{height-difference} distribution for the non-conserved EW equation in $1+1$ dimension. Finally, we discuss a UV catastrophe in the finite-time one-point distribution of height in the
(non-regularized) KPZ equation in $2+1$ dimensions.

\end{abstract}

\maketitle

\section{Introduction}

Surface growth is ubiquitous in a plethora of phenomena, from epitaxial growth to superconductors to many applications in biology \citep{Vicsek,HHZ,Barabasi,Krug}. There is a family of ``standard" stochastic growth equations which describe different classes of surface growth \citep{Barabasi,Krug}. Perhaps the most famous of them is the Kardar-Parisi-Zhang (KPZ) equation \citep{KPZ} which describes fluctuations of the height of a growing surface resulting from random deposition, surface relaxation and nonlinearity. In the KPZ equation, and in many other growth models, the interface exhibits self-affine properties, and earlier
work mostly dealt with dynamic scaling behavior of global measures such as the interface roughness \citep{Vicsek,HHZ,Barabasi,Krug}.
Recently, the focus of research on the KPZ equation shifted toward studies of the complete one-point probability distribution of the interface height at a finite time. Several groups have achieved remarkable progress in finding exact representations for this probability distribution in $1+1$ dimensions for several classes of initial conditions, see Refs. \citep{Corwin,Quastel2015,HHT,Spohn2016} for reviews.

In (much simpler) linear models, the finite-time one-point height distribution in $1+1$ dimension is well-defined for such well-known equations as the Edwards-Wilkinson (EW) equation \cite{EW1982,Krug} and the Mullins-Herring equation with conserved or non-conserved noise \cite{Krug,Mullins}. What happens in higher dimensions and/or for other surface growth models? Here we mostly address this and related questions for a class of prototypical linear stochastic growth models of the type \cite{Krug}:
\begin{equation}
\label{eq:generalized_EW}
\partial_{t}h=-\left(-\nu\nabla^{2}\right)^{m}h+\sqrt{D}\,\eta\left(\vect{x},t\right).
\end{equation}
where $h(\vect{x},t)$ is the height of the interface growing on an infinite $d$-dimensional substrate, $\nu$ is the diffusivity, $m=1,2,\dots$ is a positive integer, and $D$ is the noise magnitude. The term $\eta\left(\vect{x},t\right)$ describes a Gaussian noise with the correlation function
\begin{equation}
\label{eq:noise}
\!\!\!\!\left\langle \eta\left(\vect{x},t\right)\eta\left(\vect{x}',t'\right)\right\rangle =\left(\nabla^{2}\right)^{\alpha}\delta\left(\vect{x}-\vect{x}'\right)\delta\left(t-t'\right).
\end{equation}
For the non-conserved noise, $\alpha=0$,  $\eta$ is a white noise both in space and in time. For the conserved noise, $\alpha=1$, $\eta$ can be written as
\begin{equation}
\label{eq:noise_divergence}
\eta\left(\vect{x},t\right)=\nabla\cdot\vect{\xi}\left(\vect{x},t\right),
\end{equation}
where $\vect{\xi}$ is a white noise:
\begin{equation}
\left\langle \xi_{i}\left(\vect{x},t\right)\xi_{j}\left(\vect{x}',t'\right)\right\rangle =\delta_{ij}\delta\left(\vect{x}-\vect{x}'\right)\delta\left(t-t'\right).
\end{equation}
The systems described by Eq.~(\ref{eq:generalized_EW}) differ by their relaxation mechanism (diffusion, surface diffusion, \textit{etc}.),  character of noise (non-conserved or conserved), and the dimension of space.  For concreteness, we will assume a flat initial condition, $h\left(\vect{x},t=0\right)=0$. Because of the translational invariance of the substrate, we can study  the probability distribution   ${\mathcal{P}}\left(h_{0},t\right)$ of observing $h(\vect{x},t)=h_0$  at a finite time $t$ at \emph{any} point: for example,  at $\vect{x}=0$.

In Sec. II we will find the critical dimension, at or above which the variance of ${\mathcal{P}}\left(h_{0},t\right)$ is infinite, and so ${\mathcal{P}}\left(h_{0},t\right)$ is ill-defined unless the model is regularized at small scales. For example, for the non-conserved EW equation \citep{EW1982}, where $m=1$ and $\alpha=0$, the variance is infinite at $d\geq 2$.
For the conserved  Mullins-Herring equation ($m=2,\,\alpha=1$) \cite{Krug,Mullins} the variance of ${\mathcal{P}}\left(h_{0},t\right)$ is also infinite at $d\geq 2$, whereas
for the conserved EW equation ($m=\alpha=1$) it is infinite in \emph{all} physical dimensions.

The divergence of the finite-time variance in this class of models has the character of an ultraviolet (UV) catastrophe.
When encountering divergences like this, one usually resorts to a
microscopic cutoff for regularization \cite{Krug}. A small price to pay for such a regularization is a partial loss
of universality, as the variance now explicitly depends on the microscopic cutoff, which is different for different models
belonging to the same universality class. The difference only affects the \emph{amplitude} of the power-law
dependence of the
variance on time. Still, one can think of a simple and robust alternative that does not require a small-scale cutoff.
Here we suggest to characterize local height fluctuations by the probability distribution ${\mathcal P}[\bar{h}(t)]$ of
\emph{local average height} at time $t$, defined by averaging the surface height $h(\vect{x},t)$ over a small but
macroscopic $d$-dimensional domain $\Omega$ of volume $v$:
\begin{equation}
\label{eq:hbar_general_dim}
\bar{h}\left(t\right)=\frac{1}{v}\int_{\Omega} h\left(\vect{x},t\right)d\vect{x} .
\end{equation}
For models~(\ref{eq:generalized_EW}) the  $\bar{h}$-distribution is well-defined in arbitrary dimension.
An additional advantage of this local measure is that, at fixed time, it exhibits a crossover from a time-independent
(equilibrium or steady-state) asymptotic, obtained  for very small $v$, to a far-from-equilibrium,
time-dependent asymptotic for sufficiently large $v$. The local average height (\ref{eq:hbar_general_dim})
has been previously used for studying local roughness distributions  \citep{Halpin2014,Almeida2014,Reis2015, Carrasco2016}.
To our knowledge, the probability distribution of $\bar{h}(t)$ itself has not been previously considered,
unlike the distribution of the \emph{global} average height,
which has been studied for the KPZ equation \citep{Lee2006,Kelling2011}.

An important additional goal of this work is to show how one can use the weak-noise theory
\cite{Fogedby1998,Fogedby1999,Fogedby2009,KK2007,KK2008,MV2016,MKV_PRL2016,KMS2016,Janas2016}
to determine both the probability distribution of $\bar{h}$, and the   ``optimal path" of the interface height:  the most likely time history of  $h(\vect{x},t)$ conditioned on reaching a specified value of  $\bar{h}$ at a specified time.

As a simple test case, in Sec. III we will calculate the variance of  $\bar{h}(t)$ for the one-dimensional stochastic EW equation with conserved noise,
\begin{equation}
\label{eq:EW}
\partial_{t}h=\nu\partial_{x}^{2}h+\sqrt{D}\,\partial_{x}\xi\left(x,t\right),\quad |x|<\infty,
\end{equation}
which describes surface relaxation in the absence of deposition and desorption \citep{Kim1999}.  This is a particular case of Eq.~(\ref{eq:generalized_EW}) with $m=\alpha=d=1$, and we choose it because it is the simplest one for which the finite-time one-point distribution is ill-defined. In Sec. IV we will develop the weak-noise theory (WNT) for Eq.~(\ref{eq:generalized_EW}), and solve the WNT equations explicitly for Eq.~(\ref{eq:EW}). Sec. V deals with a closely related problem of the finite-time height-difference distribution for the \emph{non-conserved} EW equation in $1+1$ dimension. In Sec. VI we go beyond the linear models and discuss the properties of the finite-time one-point height statistics for the KPZ equation in $2+1$ dimensions. We summarize our results in Sec. VII.

\section{One-point height distribution and local average height distribution}

Consider the height-height correlation function
\begin{equation}
C\left(\vect{x}_{1},\vect{x}_{2},t\right)=\left\langle h\left(\vect{x}_{1},t\right)h\left(\vect{x}_{2},t\right)\right\rangle.
\label{corr}
\end{equation}
When  $h\left(\vect{x}_{1},t\right)$ is governed by Eq.~(\ref{eq:generalized_EW}), a standard calculation
(that we present, for completeness, in Appendix \ref{appendix:single_point}) yields
\begin{eqnarray}
\label{eq:correlation_function}
C\left(\vect{x}_{1},\vect{x}_{2},t\right)
&=& \frac{D}{\left(2\pi\right)^{d}}\int d\vect{k} \, e^{i\vect{k}\cdot\left(\vect{x}_{1}-\vect{x}_{2}\right)} \nonumber\\
&\times& \frac{k^{2\alpha-2m}}{2\nu^{m}}\left[1-e^{-2\left(k^{2}\nu\right)^{m}t}\right].
\end{eqnarray}
As one can see,  $C\left(\vect{x_{1}}\neq \vect{x_{2}},t>0\right)$ is well-defined because the integral over
$\mathbf{k}$ converges. To show it, we can set $\vect{x_{1}} = 0$, because the system is homogeneous in space.
The correlator $C\left(0,\vect{x},t\right)$ can depend only on the distance $x = \left|\vect{x}\right|$,
because the system is isotropic. Correspondingly, it is convenient to evaluate the integral
(\ref{eq:correlation_function}) in the ($d$-dimensional) spherical coordinates. Integrations
over all the angles give a function of $x$, and only a single integral, over $k= \left|\vect{k}\right|$,
remains. At $\vect{x}\neq 0$ this integral converges at $k\to \infty$ due to the oscillatory term
$e^{i\vect{k}\cdot\vect{x}}$ in the original integrand. The convergence at $k=0$ is guaranteed,
at finite $t$,  by the time-dependent factor inside the square brackets under the integral.

Now we can address the finite-time interface-height variance at a point $\vect{x}$. This quantity is immediately obtained
from $C\left(\vect{x}_{1},\vect{x}_{2},t\right)$:
\begin{equation}
\label{eq:variance_def}
\!\! \text{Var}\left[h\left(\vect{x},t\right)\right]=
\left\langle h\left(\vect{x},t\right)^{2}\right\rangle =C\left(\vect{x},\vect{x},t\right)=C\left(0,0,t\right);
\end{equation}
it is independent of $\vect{x}$. It is easily seen from Eq.~(\ref{eq:correlation_function}) that, at finite $t$, the variance (\ref{eq:variance_def}) is finite if and only if the space dimension is smaller than the critical dimension:
\begin{equation}
\label{eq:critical_dimension}
d<d_{c}=2m-2\alpha.
\end{equation}
For $d \ge d_c$, the variance (\ref{eq:variance_def}) diverges at $k\to \infty$. This divergence -- a UV
catastrophe -- is present both in infinite and in finite systems.
For example, for the EW equation with non-conserved noise ($m=1, \alpha=0$) the critical dimension
(\ref{eq:critical_dimension}) is $d_c=2$, while for conserved noise ($\alpha=1$) it is $d_c=0$. For
the Mullins-Herring equation with non-conserved noise ($m=2, \alpha=0$), the critical dimension is $d_c=4$,
while for conserved noise ($\alpha=1$) it is $d_c=2$.

The UV catastrophe at $d \ge d_c$ is not unique to the finite-time one-point height distribution. In finite systems, describable by Eqs.~(\ref{eq:generalized_EW}) and (\ref{eq:noise}) in the absence of a small-scale regularization,  the finite-time interface width
\begin{equation}
\label{eq:widthglobal}
W=\left\langle \frac{1}{V}\int d\vect{x_{1}}\,\left[h\left(\vect{x_{1}},t\right)-\frac{1}{V}\int d\vect{x_{2}}h\left(\vect{x_{2}},t\right)\right]^{2}\right\rangle ^{1/2}
\end{equation}
(where the spatial integration is over the entire system, and $V$ is the system's volume)
also diverges, at $d \ge d_c$,  due to the divergence of the term $\left\langle h\left(\vect{x_{1}},t\right)^{2}\right\rangle$. In this context the
UV catastrophe
is well known to experts. For example, for the non-conserved noise, when $d_c=2m$,
it is evident from Eq.~(3.28) of the review \citep{Krug}.

In practice, the UV catastrophe is usually avoided by introducing a small-scale cutoff such
as the lattice constant, finite correlation length of the noise, \textit{etc}. This leads to a partial
loss of universality, as explained in the Introduction. A finite correlation length can also cause difficulties in
attempts of exact solution.  As a possible alternative that keeps the noise white in space, we suggest to characterize local fluctuations of the interface by the distribution of the local average height
(\ref{eq:hbar_general_dim}). Let us assume, for concreteness, that the spatial average in Eq.~(\ref{eq:hbar_general_dim})
is performed over a $d$-dimensional hypercube $\left[-L,L\right]^{d}$. Since Eqs.~(\ref{eq:generalized_EW}) and
(\ref{eq:hbar_general_dim}) are linear in $h$, the fluctuations of $\bar{h}$ are Gaussian, and it suffices to
evaluate their variance. See Appendix \ref{appendix:local_average_height} for a brief derivation. The result is
\begin{eqnarray}
\label{eq:variance_exact}
\text{Var}\left[\bar{h}\left(t\right)\right] &=&
\frac{D}{\left(2\pi\right)^{d}L^{2d}}\int d\vect{k}\,\prod_{i=1}^{d}\frac{\sin^{2}\left(k_{i}L\right)}{k_{i}^{2}} \nonumber\\
& \times & \frac{k^{2\alpha-2m}}{2\nu^{m}}\left[1-e^{-2\left(k^{2}\nu\right)^{m}t}\right].
\end{eqnarray}
It is straightforward to show that the integral in Eq.~(\ref{eq:variance_exact}) converges in all dimensions,
so for the models~(\ref{eq:generalized_EW}) and (\ref{eq:noise}) this quantity is well-defined.

\section{EW equation with conserved noise}
\label{sec:EW_fourier}
As a simple illustration, we consider Eq.~(\ref{eq:EW}). Formally, it can be viewed as a particular case of the Langevin equation
\begin{equation}
\partial_{t}\rho=\nabla\cdot\left[{\mathcal D}\left(\rho\right)\nabla\rho+\sqrt{\sigma\left(\rho\right)} \, \xi\right]
\end{equation}
which provides a coarse-grained description of a family of diffusive lattice gases with density $\rho\left(x,t\right)$, diffusivity   ${\mathcal D}\left(\rho\right)$ and mobility $\sigma\left(\rho\right)$ \cite{Spohn1991}. In this particular case the diffusivity and mobility
of the ``lattice gas" are both density-independent.
For diffusive lattice gases, the equilibrium  can be described in terms of a free energy density $\mathcal{F}\left(\rho\right)$ which satisfies the fluctuation-dissipation relation \citep{Spohn1991, Derrida2007}
\begin{equation}
\mathcal{F}^{\prime\prime}\left(\rho\right)=\frac{2{\mathcal D}\left(\rho\right)}{\sigma\left(\rho\right)}.
\end{equation}
For  Eq.~(\ref{eq:EW}) this gives
\begin{equation}
\label{eq:free_energy}
\mathcal{F}\left(h\right)=\frac{\nu h^{2}}{D}.
\end{equation}
For $d=1$, and in the limit of $t\to\infty$, Eq.~(\ref{eq:correlation_function}) yields:
\begin{equation}
C(x_1,x_2,t\to \infty) =\frac{D}{2 \nu}\delta\left(x_{1}-x_{2}\right).
\label{CEWcorrelator}
\end{equation}
Indeed, the interface height at thermal equilibrium  is delta-correlated, which is consistent with the UV catastrophe of the one-point height variance. Equation~(\ref{CEWcorrelator}) also directly follows from Eq.~(\ref{eq:free_energy}) \cite{Spohn1991}.

According to Eq.~(\ref{eq:critical_dimension}), the critical dimension for this model is zero, so the finite-time one-point height distribution of this model is ill-defined in all physical dimensions. Let us determine the distribution of the local average height (\ref{eq:hbar_general_dim}). Rescale time $t$ by the observation time $T$, the spatial coordinate $x$ by $\sqrt{\nu T}$ and the interface height $h$ by $\kappa=D^{1/2}\nu^{-3/4}T^{-1/4}$.  The resulting rescaled conserved EW equation is parameter-free:
\begin{equation}
\label{eq:langevin_dimensionless}
\partial_{t}h=\partial_{x}^{2}h+\partial_{x}\xi\left(x,t\right).
 \end{equation}
The local average height (\ref{eq:hbar_general_dim}) at $t=1$, in the rescaled variables, is
\begin{equation}
\label{eq:qbar_def}
\bar{h}\left(t=1\right)=\frac{1}{2\ell}\int_{-\ell}^{\ell}h\left(x,1\right)dx,
 \end{equation}
where $\ell=L/\sqrt{\nu T}$. Equation~(\ref{eq:variance_exact}) yields
\begin{equation}
\label{eq:variance_dimnesionless_exact}
\text{Var}\left[\bar{h}\left(t=1\right)\right]=\frac{1}{4\ell^{2}}\left[\sqrt{\frac{2}{\pi}}\left(1-e^{-\frac{\ell^{2}}{2}}\right)+\ell \, \text{erfc}\left(\frac{\ell}{\sqrt{2}}\right)\right],
\end{equation}
where $\text{erfc} \,z = 1-\text{erf}\,z = (2/\sqrt{\pi}) \int_z^{\infty} e^{-\zeta^2}\,d\zeta$.
In the physical units,
\begin{eqnarray}
\label{eq:variance_physical_units}
\text{Var}\left[\bar{h}\left(t=T\right)\right] &=& \frac{D}{4L^{2}}\sqrt{\frac{T}{\nu}}\left[\sqrt{\frac{2}{\pi}}\left(1-e^{-\frac{L^{2}}{2\nu T}}\right)\right. \nonumber\\
&+&\left.\frac{L}{\sqrt{\nu T}} \, \text{erfc}\left(\frac{L}{\sqrt{2\nu T}}\right)\right].
\end{eqnarray}
Because of the term including $\text{erfc}$, these expressions diverge at $\ell\to 0$, or $L\to 0$, as expected. We now examine the long- and short-time behaviors of the variance.
In the long-time limit, $\ell \ll 1$, the leading-order asymptote is
\begin{equation}
\label{eq:variance_small_l_limit}
\text{Var}\left[\bar{h}\left(t=1\right)\right]\simeq \frac{1}{4\ell} ,
 \end{equation}
Correspondingly, the local average height distribution is
\begin{equation}
\label{eq:distribution_small_l_limit}
P\left[\bar{h}\left(t=1\right)\right]\simeq\sqrt{\frac{2\ell}{\pi}} \, e^{-2\ell\bar{h}\left(1\right)^{2}}.
\end{equation}
In the physical variables, the distribution is
\begin{equation}\label{eq:distribtion_small_l_phys}
{\mathcal P}\left[\bar{h}\left(t=T\right)\right]\simeq \left(\frac{2\nu L}{\pi D}\right)^{1/2} \, e^{-\frac{2\nu L \bar{h}^2}{D}};
\end{equation}
it is independent of $T$ as expected from an equilibrium distribution. Furthermore, if we assume that
\begin{equation}
h\left(x,t=T\right)\simeq\begin{cases}
\bar{h}, & \left|x\right|<L,\\
0, & \left|x\right|>L,
\end{cases}
\label{table}
\end{equation}
 then the term
$$
\frac{2 \nu L \bar{h}\left(t=T\right)^{2}}{D} = \frac{\nu \bar{h}\left(t=T\right)^{2}}{D} \times  2L
$$
describes the increase of the free energy of the interface compared with the flat state $h=0$, see Eq.~(\ref{eq:free_energy}). The optimal interface history, that we will determine shortly, fully supports this interpretation.

Accounting for the subleading correction to Eq.~(\ref{eq:variance_small_l_limit}), we obtain
\begin{equation}
\text{Var}\left[\bar{h}\left(1\right)\right]\simeq\frac{1}{4\ell}\left(1-\frac{\ell}{\sqrt{2\pi}}\right).
\end{equation}
The correction is negative, so the probability to observe the same $\bar{h}(t=1)$ is smaller than in equilibrium,
as to be expected on the physical grounds.

In the short-time limit, $\ell \gg 1$, Eq.~(\ref{eq:variance_dimnesionless_exact}) becomes
\begin{equation}
\label{eq:variance_large_l_limit}
\text{Var}\left[\bar{h}\left(t=1\right)\right]\simeq\frac{1}{2\sqrt{2\pi} \, \ell^{2}}.
\end{equation}
The probability to observe a given $\bar{h}$ at short times is strongly suppressed as to be expected. To better understand Eq.~(\ref{eq:variance_large_l_limit}), let us calculate the variance of the total rescaled ``mass", $2\ell\bar{h}\left(1\right)$, which enters the interval $\left[-\ell,\ell\right]$ for $\bar{h}>0$, or exits this interval for $\bar{h}<0$:
\begin{equation}
\label{varnoell}
\text{Var}\left[2\ell\bar{h}\left(t=1\right)\right] = 4\ell^{2}\text{Var}\left[\bar{h}\left(t=1\right)\right]\simeq \sqrt{\frac{2}{\pi}}.
\end{equation}
This quantity does not depend on $\ell$, and the reason for this will become clear when we determine the optimal interface history in this limit.

\section{Optimal interface history}

\subsection{General}

Now we return to the more general Eq.~(\ref{eq:generalized_EW}) and show how one can use the weak-noise theory (WNT) of stochastic surface growth \citep{Fogedby1998,Fogedby1999,Fogedby2009,KK2007,KK2008,MV2016,MKV_PRL2016,KMS2016,Janas2016}
to determine the optimal history $h\left(\vect{x},t\right)$ of the interface profile, conditioned on a given value of $\bar{h}$ at a specified time $t=T$. For nonlinear evolution equations, the leading-order calculations of the WNT theory enable  one to evaluate the distribution of $\bar{h}(T)$  only up to pre-exponential factors. For linear equations, like Eq.~(\ref{eq:generalized_EW}), the expected distributions are Gaussian. Therefore, the pre-exponential factors can be found from normalization, and the WNT yields exact results.
The WNT equations can be obtained via a  saddle-point evaluation of the action integral for Eq.~(\ref{eq:generalized_EW}), see Appendix \ref{appendix:WNT}. They can be written as Hamilton's equations for the optimal path $h(x,t)$ and the conjugate ``momentum density" $p(x,t)$:
\begin{eqnarray}
  \partial_{t}h &=& \delta H/\delta p=-\left(-\nu\nabla^{2}\right)^{m}h+\left(-\nabla^{2}\right)^{\alpha}p, \label{eq:q_general_dim} \\
  \partial_{t}p &=&  -\delta H/\delta h=\left(-\nu\nabla^{2}\right)^{m}p, \label{eq:p_general_dim}
\end{eqnarray}
where the Hamiltonian is
\begin{equation}\label{H1general}
\!\!H=\int_{-\infty}^{\infty}\!\!\!\!dx\,\mathcal{H},\quad\mathcal{H}=
-h\left(-\nu\nabla^{2}\right)^{m}p+\frac{1}{2}\left(\nabla^{\alpha}p\right)^{2}.
\end{equation}
The initial condition for the flat interface is
\begin{equation}
h\left(\vect{x},t=0\right)=0.
\end{equation}
An additional condition, at $t=T$, stems from the integral constraint (\ref{eq:hbar_general_dim}). As shown in Appendix \ref{appendix:WNT}, it has the form
\begin{equation}
\label{eq:initial_condition_p_general}
p\left(\vect{x},t=T\right)=\begin{cases}
\Lambda, & \vect{x}\in\Omega ,\\
0, & \vect{x}\notin\Omega ,
\end{cases}
\end{equation}
where $\Lambda$ is an a priori unknown Lagrange multiplier whose value is ultimately set by Eq.~(\ref{eq:hbar_general_dim}).
Once Eqs.~(\ref{eq:q_general_dim}) and (\ref{eq:p_general_dim}) are solved, one can evaluate the probability
$\mathcal{P}$ of observing a specified value of $\bar{h}(T)$ from the relation $-\ln{\mathcal{P}}\simeq S/D$,
where
\begin{equation}
S=\frac{1}{2}\int_{0}^{T}dt\int d\vect{x}\,\left(\nabla^{\alpha}p\right)^{2}
\end{equation}
is the action evaluated along the optimal path.

For the KPZ equation in one dimension (which includes, as a simple limit, the non-conserved EW equation), the WNT was employed in Refs. \cite{KK2007,MKV_PRL2016,KMS2016,Janas2016} for determining the finite-time one-point height distribution for different initial conditions. The WNT was also used for the stochastic Mullins-Herring equation with conserved and non-conserved noise \citep{MV2016}. We now proceed to solve the WNT equations for the one-dimensional conserved EW equation when the process is conditioned on a given local average height. As the conserved EW equation can be formally viewed as a lattice gas, here the WNT equations represent a particular case of the macroscopic fluctuation theory  of lattice gases \citep{Bertini2015}.

\subsection{Conserved EW Equation in 1+1 Dimensions}

Upon the rescaling of $x$, $t$ and $h$  leading to Eq.~(\ref{eq:langevin_dimensionless}), Eqs.~(\ref{eq:q_general_dim}) and (\ref{eq:p_general_dim}) become
\begin{eqnarray}
  \partial_{t}h &=& \delta H/\delta p  \; = \partial_{x}^{2}h - \partial_{x}^{2}p, \label{eq:q} \\
   \partial_{t}p &=& \! -\delta H/\delta h= -\partial_{x}^{2}p,\label{eq:p}
\end{eqnarray}
where $p$ is rescaled by $\nu\kappa=D^{1/2}\nu^{1/4}T^{-1/4}$.
The rescaled Hamiltonian is
\begin{equation}\label{H1}
H=\int_{-\infty}^{\infty} \!\!\! dx\,\mathcal{H},\quad\mathcal{H}=\partial_{x}p\left(-\partial_{x}h+\frac{1}{2}\partial_{x}p\right),
\end{equation}
while the rescaled action, $s=S/D$, is
\begin{equation}
\label{action}
s=\int_{0}^{1} \! dt\int_{-\infty}^{\infty}  \!\!\! dx\,\left(p\partial_{t}h-\mathcal{H}\right)=\frac{1}{2}\int_{0}^{1} \! dt\int_{-\infty}^{\infty} \!\!\! dx\,\left(\partial_{x}p\right)^{2} .
\end{equation}
Integrating by parts and using Eq.~(\ref{eq:p}), we obtain a convenient expression for $s$ which does not involve integration over time:
\begin{eqnarray}
\label{eq:action_space_integral}
s \!\!&=& \! -\frac{1}{2}\int_{0}^{1} \!\! dt\int_{-\infty}^{\infty} \!\!\!\! dx\,p\,\partial_{x}^{2}p=\frac{1}{2}\int_{-\infty}^{\infty} \!\!\!\! dx\int_{0}^{1} \!\! dt\,p\,\partial_{t}p  \nonumber\\
&=& \! \frac{1}{4}  \! \int_{-\infty}^{\infty} \!\!\!\!\! dx \! \int_{0}^{1} \!\!\! dt\,\partial_{t}\left(p^{2}\right) = \frac{1}{4} \! \int_{-\infty}^{\infty} \!\!\!\!\! dx\left[p^{2}\left(x,1\right)-p^{2}\left(x,0\right)\right].\nonumber\\
\end{eqnarray}
As follows from Eq.~(\ref{H1}), there are two invariant zero-energy manifolds. The manifold $\partial_x p=0$ corresponds to
the deterministic EW equation $\partial_t h = \partial_x^2 h$.
The second zero-energy manifold,
\begin{equation}\label{eqmanifold}
p = 2h ,
\end{equation}
describes thermal equilibrium. Indeed, Eqs.~(\ref{eq:q}) and (\ref{eqmanifold}) yield the time-reversed deterministic
EW equation
\begin{equation}\label{timereversed}
\partial_t h = -\partial_x^2 h.
\end{equation}
Therefore, an activation trajectory at equilibrium coincides with the time-reversed relaxation trajectory, as to  be expected \cite{Onsager}. In the limit $\ell \ll 1$,  the system has sufficient time to explore equilibrium fluctuations in order to reach the specified local average height.  In this limit the action must be equal to the difference between the free energies of the final and initial states. Indeed, evaluating the action, using Eq.~(\ref{eq:action_space_integral}), on the equilibrium manifold~(\ref{eqmanifold}), we obtain
\begin{equation}
\label{eq:action_thermal_equilibrium}
s=\int_{-\infty}^{\infty} \!\!\! dx \, h^{2}\left(x,t=1\right),
\end{equation}
the (rescaled) free energy (\ref{eq:free_energy}) cost of the height profile $h\left(x,t=1\right)$. This cost must be minimized with respect to all possible height profiles $h\left(x,t=1\right)$ with local average height $\bar{h}$ (\ref{eq:qbar_def}). As a result, the minimum is achieved on a discontinuous height profile:
\begin{equation}
\label{eq:optimal_qfinal_in_equilibrium}
h\left(x,t=1\right)=\begin{cases}
\bar{h}, & \left|x\right|<\ell,\\
0, & \left|x\right|>\ell,
\end{cases}
\end{equation}
and so $s=2\ell\bar{h}^{2}$, in agreement with Eq.~(\ref{eq:distribution_small_l_limit}).

For finite $\ell$ the system does not live on the equilibrium manifold, and we must solve Eqs.~(\ref{eq:q}) and (\ref{eq:p}) explicitly,
with boundary conditions $h\left(x,t=0\right)=0$
and
\begin{equation}
p\left(x,t=1\right)=\begin{cases}
\lambda, & \left|x\right|<\ell,\\
0, & \left|x\right|>\ell,
\end{cases}
\label{p1}
\end{equation}
with an a priori unknown $\lambda$. Solving  Eq.~(\ref{eq:p}) backward in time with the initial condition (\ref{p1}), we obtain
\begin{equation}
\label{eq:sol_p}
\!\! p\left(x,t\right) \! = \! \frac{\lambda}{2} \! \left[\text{erf}\left( \! \frac{x+\ell}{\sqrt{4\left(1-t\right)}}\right) \! -\text{erf}\left( \! \frac{x-\ell}{\sqrt{4\left(1-t\right)}}\right) \! \right] \! .
\end{equation}
Next, we introduce the auxiliary field
\begin{equation}
\label{eq:r_def}
r\left(x,t\right)\equiv h\left(x,t\right) - \frac{1}{2}p\left(x,t\right).
\end{equation}
Using Eqs.~(\ref{eq:q})--(\ref{eq:p}), one can see that this field satisfies the diffusion equation
\begin{equation}
\label{eq:r}
\partial_{t}r=\partial_{x}^{2}r.
\end{equation}
Using the flat initial condition for $h$ and evaluating $p\left(x,t=0\right)$ from Eq.~(\ref{eq:sol_p}), we can solve Eq.~(\ref{eq:r}):
\begin{equation}
\label{eq:sol_r}
\!\!\! r \! \left(x,t\right) \! = \! - \frac{\lambda}{4} \! \left[\text{erf}\left( \! \frac{x+\ell}{\sqrt{4\left(1+t\right)}}\right) \!\! - \! \text{erf}\left( \! \frac{x-\ell}{\sqrt{4\left(1+t\right)}}\right) \! \right] \! .
\end{equation}
Plugging Eqs.~(\ref{eq:sol_p}) and (\ref{eq:sol_r}) into Eq.~(\ref{eq:r_def}), we obtain the optimal interface height profile:
\begin{eqnarray}
\label{eq:sol_q}
h\left(x,t\right) \! &=& \! - \frac{\lambda}{4}\left[\text{erf}\left(\frac{x-\ell}{\sqrt{4\left(1-t\right)}}\right)-
\text{erf}\left(\frac{x+\ell}{\sqrt{4\left(1-t\right)}}\right)\right.\nonumber\\
&+& \! \left.\text{erf}\left(\frac{x+\ell}{\sqrt{4\left(1+t\right)}}\right)-
\text{erf}\left(\frac{x-\ell}{\sqrt{4\left(1+t\right)}}\right)\right].
\end{eqnarray}
The Lagrange multiplier $\lambda$ is then found from Eq.~(\ref{eq:qbar_def}):
\begin{equation}
\label{sol_lambda}
\lambda=\frac{2\ell\bar{h}}{\ell+\sqrt{2/\pi} \, \left(1-e^{-\ell^{2}/2}\right)-\ell \, \text{erf}\left(\ell/\sqrt{2}\right)} \, .
\end{equation}
\begin{figure}
\includegraphics[width=0.4\textwidth,clip=]{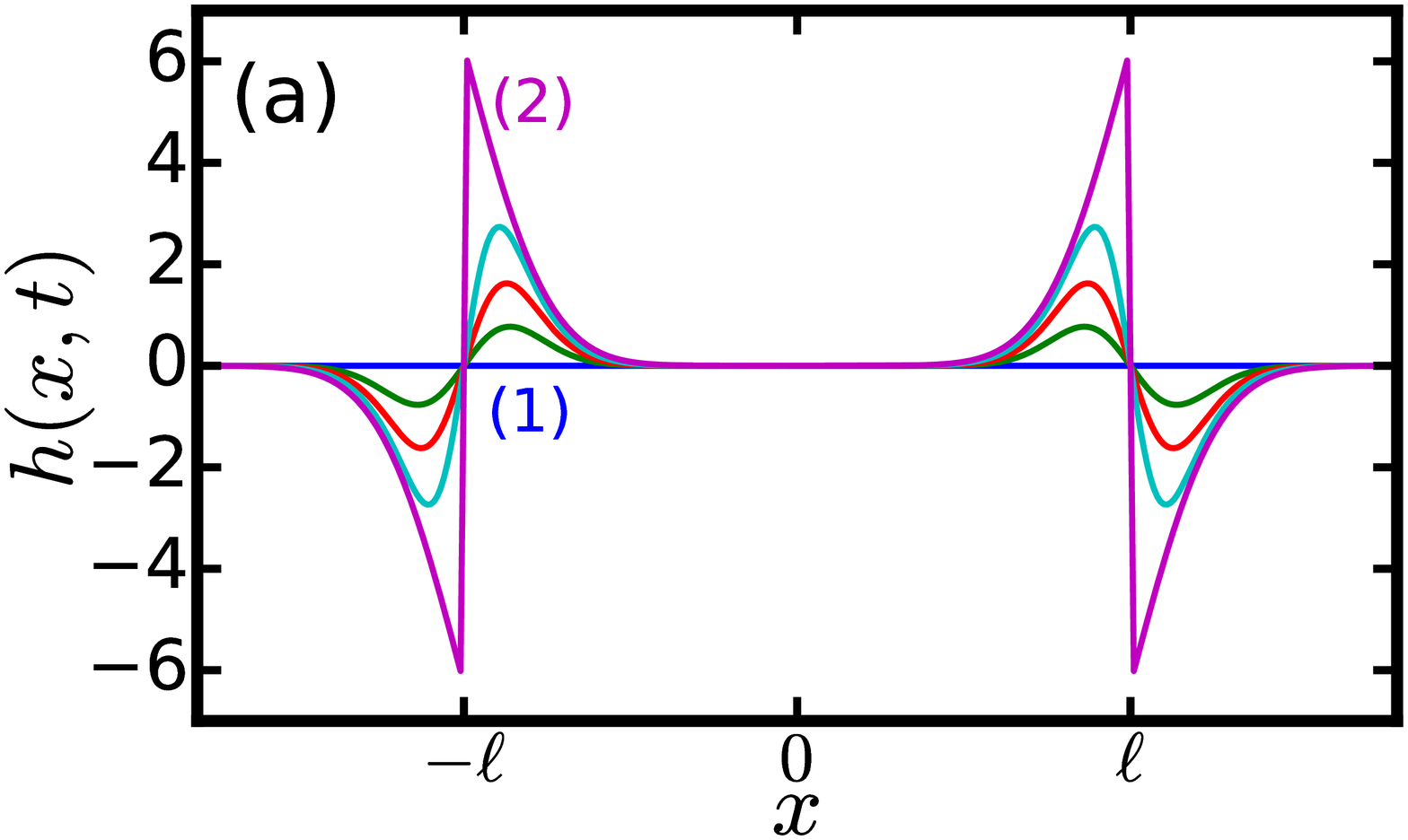}
\includegraphics[width=0.4\textwidth,clip=]{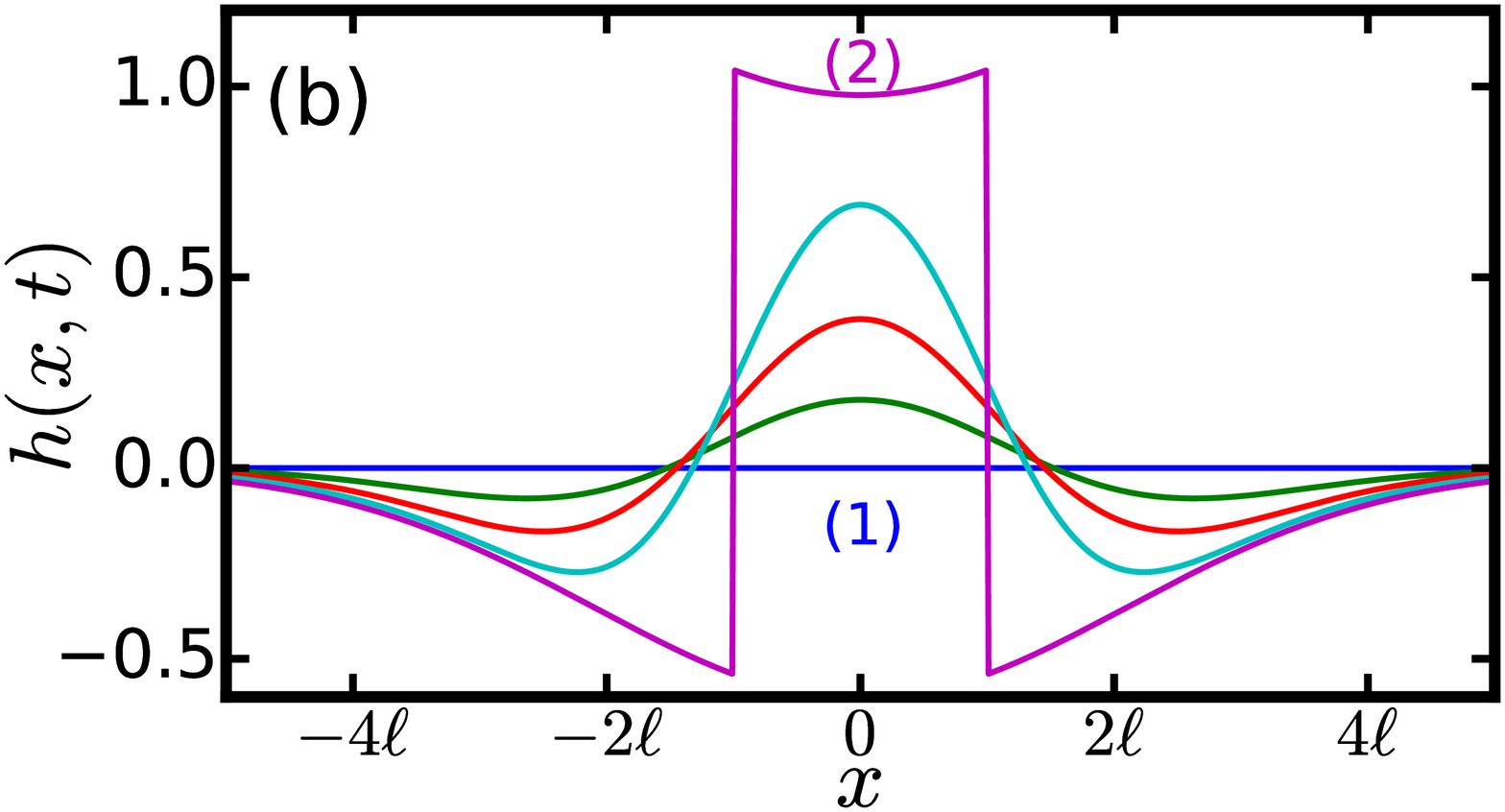}
\includegraphics[width=0.4\textwidth,clip=]{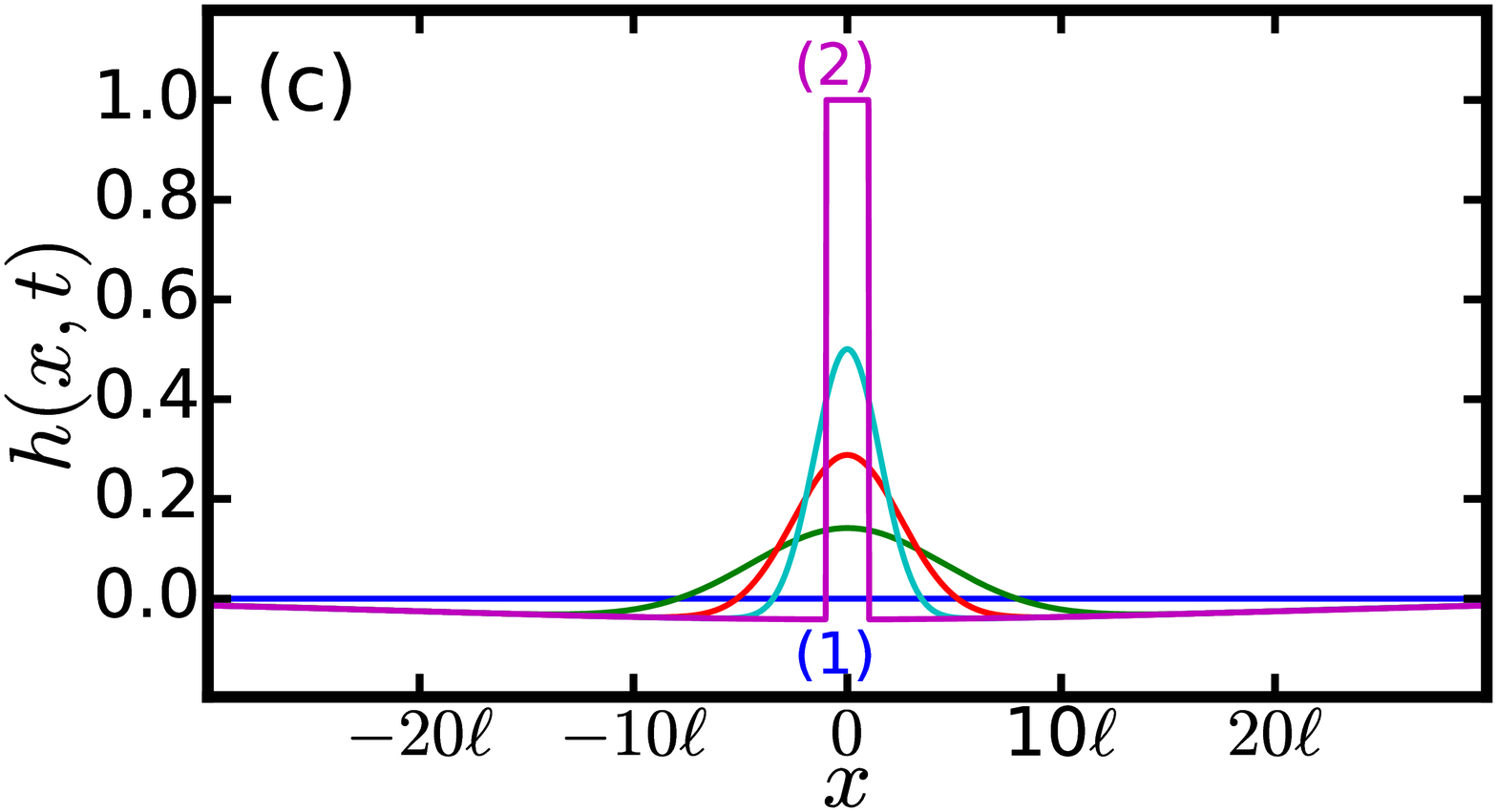}
\caption{The optimal path of the interface, conditioned on reaching a rescaled local average height $\bar{h}=1$ at $t=T$. Shown is the height $h(x,t)$, rescaled by $\kappa=D^{1/2}\nu^{-3/4}T^{-1/4}$, as a function of the rescaled coordinate $x/\sqrt{\nu T}$, for $\ell = 10$ (a), $\ell = 1$ (b) and $\ell = 0.1$ (c). The initial ($t=0$) and final ($t=T$) profiles are marked by (1) and (2), respectively. On panels (a) and (b) the height is plotted at times $t=0$, $T/4$, $T/2$, $3T/4$, and $T$. On panel (c) the times are $t=0$, $0.9T$, $0.97T$, $0.99T$, and $T$.}
\label{fig:optimal_history}
\end{figure}
Figure \ref{fig:optimal_history} shows the optimal paths $h(x,t)$ for $\ell=10$, $1$ and $0.1$. For large $\ell$, or short times,  the optimal interface dynamics are localized at
the boundaries $x=\pm \ell$. This is the reason that $\ell$ drops from Eq.~(\ref{varnoell}). For $\ell \to 0$ the optimal profile at $t=T$ approaches the equilibrium one, described by Eq.~(\ref{eq:optimal_qfinal_in_equilibrium}). Furthermore, the profile stays flat in this case, $h(x,t)\simeq 0$, for most of the dynamics, and the optimal fluctuation only develops towards the end.
For very small $\ell$ in Eqs.~(\ref{eq:sol_q}) and (\ref{sol_lambda}), the final interface profile is approximated by
\begin{equation}
\label{eq:final_profile_approx}
\! h\left(x,1\right)\simeq\bar{h}\left[\theta\left(x-\ell\right)-\theta\left(x+\ell\right)+ \! \frac{\ell}{\sqrt{2\pi}}\exp\left( \! -\frac{x^{2}}{8}\right) \! \right].
\end{equation}
The first two terms in Eq.~(\ref{eq:final_profile_approx}) are the thermal equilibrium terms, while the last term is the leading non-equilibrium correction.

\begin{figure}
\includegraphics[width=0.42\textwidth,clip=]{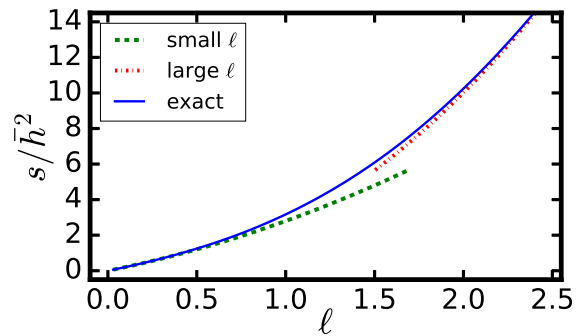}
\caption{The rescaled action divided by the local average height squared (solid line), alongside with the small-$\ell$ limit (dashed) and the large-$\ell$ limit (dot-dashed).}
\label{fig:action}
\end{figure}

We now turn to the evaluation of the rescaled action (\ref{action}).
Using Eqs.~(\ref{eq:sol_p}) and (\ref{sol_lambda}) in Eq.~(\ref{eq:action_space_integral}), and performing the integral over $x$ (see Appendix \ref{appendix:erf_integral}), we obtain
\begin{equation}
\label{eq:S_sol}
s=\frac{2\ell^{2}\bar{h}^{2}}{\ell\,\text{erfc}\left(\ell/\sqrt{2}\right)+\sqrt{2/\pi}\,\left(1-e^{-\ell^{2}/2}\right)} \, .
\end{equation}
As expected, this corresponds to a Gaussian distribution ${\mathcal P}(\bar{h})$ whose variance, $\bar{h}^2/\left(2s\right)$, coincides with the exact result (\ref{eq:variance_dimnesionless_exact}). In the physical units, the action $S=sD$ coincides with the result from Eq.~(\ref{eq:variance_physical_units}).
A graph of $s/\bar{h}^2$ as a function of $\ell$ is plotted in Fig.~\ref{fig:action}.

It is interesting that, similarly to other non-equilibrium problems that are exactly soluble in the framework of a weak-noise theory \cite{KMSvoid}, one can define the ``non-equilibrium free energy"
\begin{equation}
F\left[p\left(x\right)\right]=\frac{1}{4}\int_{-\infty}^{\infty}\!\!\!dx\,p^{2}\left(x\right),
\end{equation}
so that the dimensionless action (\ref{eq:action_space_integral}) is the difference between the values of $F$ at $t=1$ and $t=0$. This simplification, however, does not save us the need of solving  the dynamical problem in order to evaluate the momentum density $p\left(x,t\right)$ at $t=1$ and $t=0$.

\section{EW equation with non-conserved noise}

\subsection{Distribution of the Height Difference}

Now let us consider the nonconserved EW equation \citep{EW1982}
\begin{equation}
\label{eq:EW_nc}
\partial_{t}h=\nu\partial_{x}^{2}h+\sqrt{D_0}\,\xi\left(x,t\right).
\end{equation}
Taking the derivative of Eq.~(\ref{eq:EW_nc}) gives us the equation
\begin{equation}
\label{eq:EW_f}
\partial_{t}f=\nu\partial_{x}^{2}f+\sqrt{D_0}\, \partial_x \xi\left(x,t\right),
\end{equation}
where $f=\partial h/\partial x$. Eq.~(\ref{eq:EW_f}) is mathematically equivalent to Eq.~(\ref{eq:EW}). Furthermore, there is a simple connection between the local average of $f$ and the height difference function of $h$:
\begin{equation}
\bar{f}\left(t\right)=\frac{1}{2L}\int_{-L}^{L}f\left(x,t\right)dx=\frac{1}{2L}\int_{-L}^{L}\frac{\partial h\left(x,t\right)}{\partial x}dx=\frac{\Delta}{2L},
\end{equation}
where $\Delta=h\left(L,t\right)-h\left(-L,t\right)$.
The probability distribution of $\Delta$ at a specified time $T$ is therefore immediately obtained from the distribution of $\bar{f}$; it is a Gaussian distribution whose variance is given by
\begin{equation}
\text{Var}\left[\Delta\left(T\right)\right]=4L^{2}\text{Var}\left[\bar{f}\left(T\right)\right].
\end{equation}
Using Eq.~(\ref{eq:variance_physical_units}), we obtain:
\begin{equation}
\label{eq:Var_Delta_nonconserving}
\text{Var}\!\left(\Delta\right)\!=\!D_{0}\sqrt{\frac{T}{\nu}}\!\left[\!\sqrt{\!\frac{2}{\pi}}\!\left(1\!-e^{-\frac{L^{2}}{2\nu T}}\right)\!+\!\frac{L}{\sqrt{\nu T}}\,\text{erfc}\!\left(\!\frac{L}{\sqrt{2\nu T}}\right)\!\right]\!.
\end{equation}
This result has been known for a long time, see Eqs.~(2.19) and (2.21) of Ref.~\cite{Natterman1992,Footnote2}. As shown below, the WNT gives an additional insight into the problem by providing us with the optimal path of the system conditioned on a specified height difference.

\subsection{Optimal History Conditioned on Height Difference}
\begin{figure}[ht]
\includegraphics[width=0.4\textwidth,clip=]{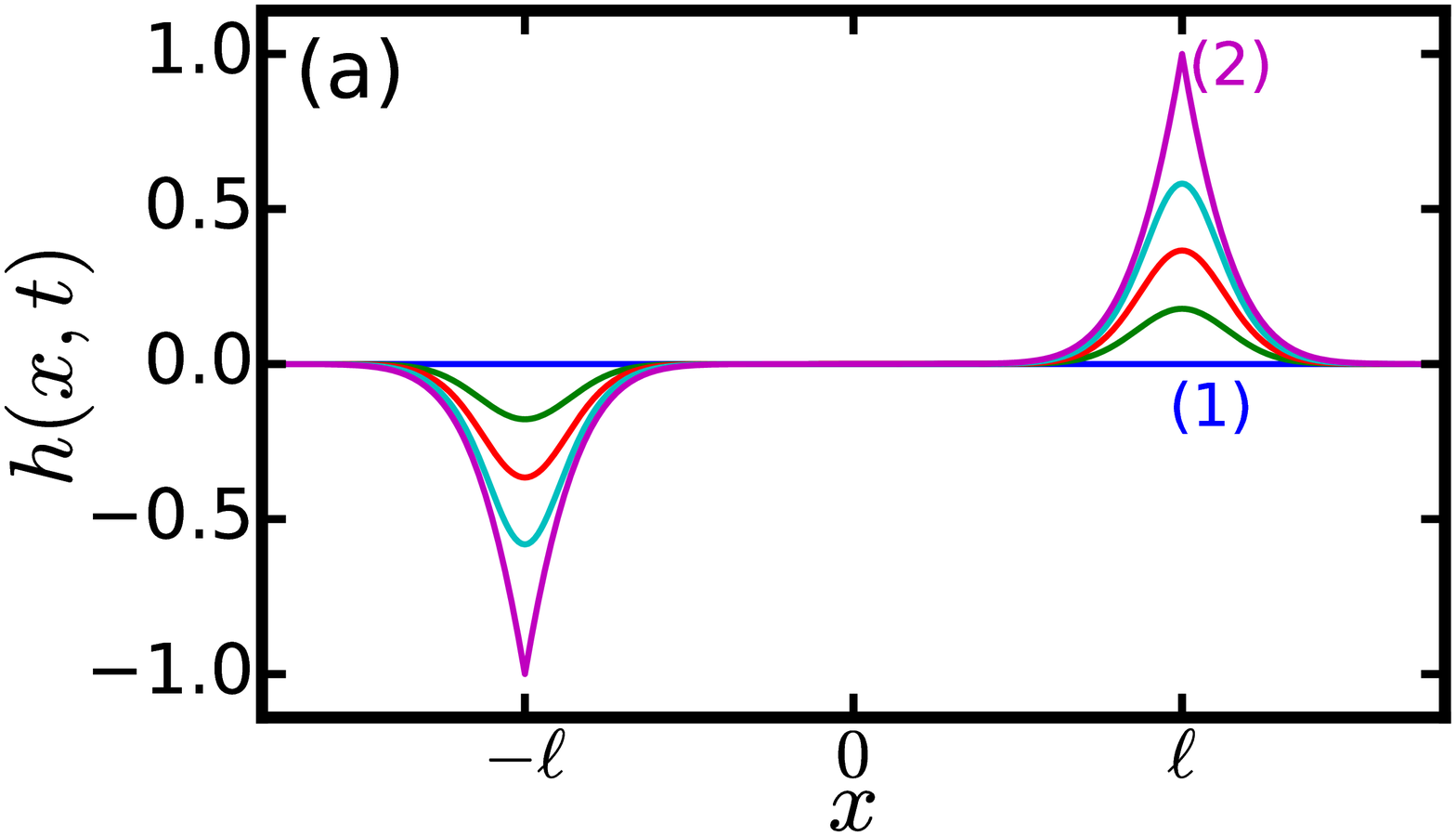}
\includegraphics[width=0.4\textwidth,clip=]{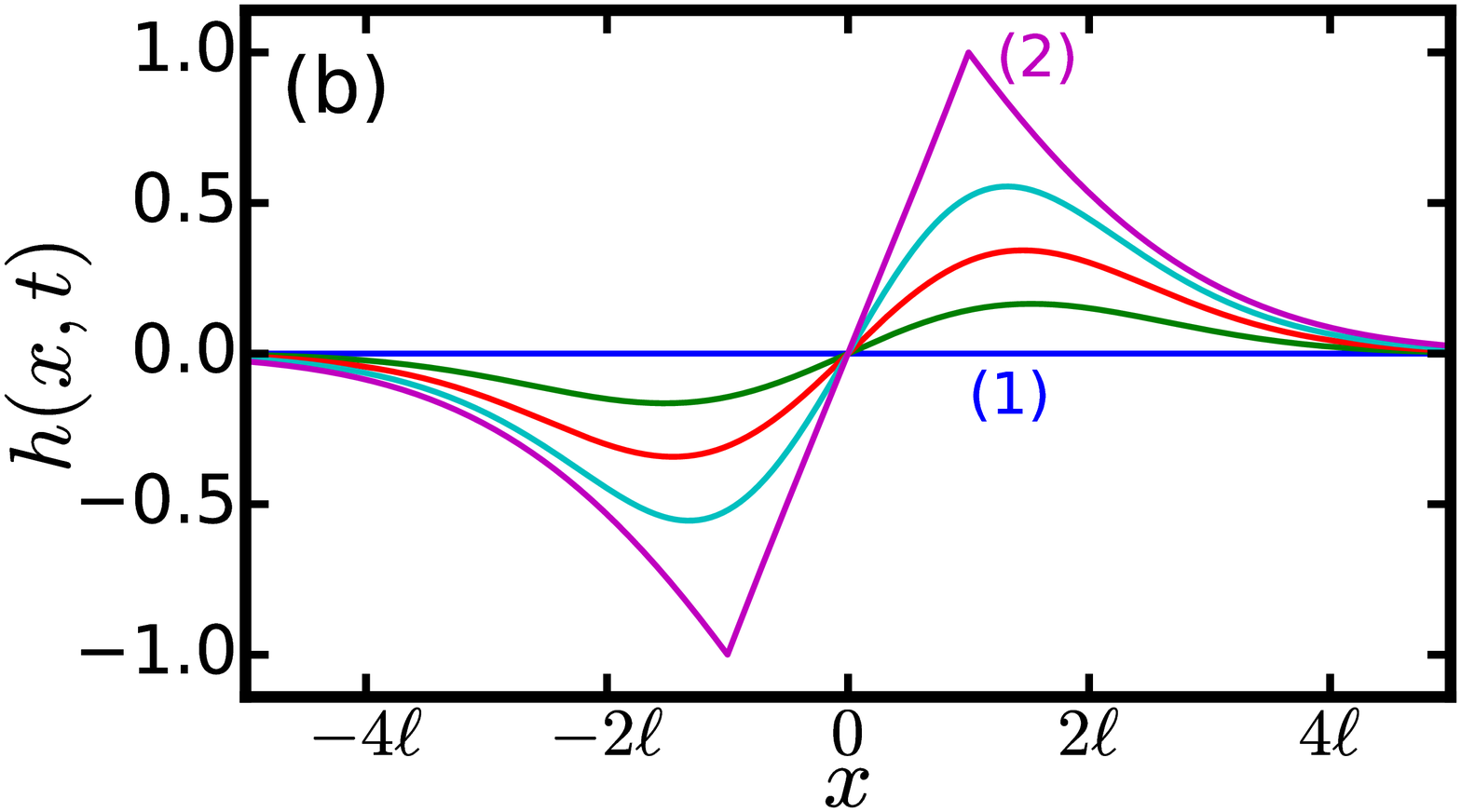}
\includegraphics[width=0.4\textwidth,clip=]{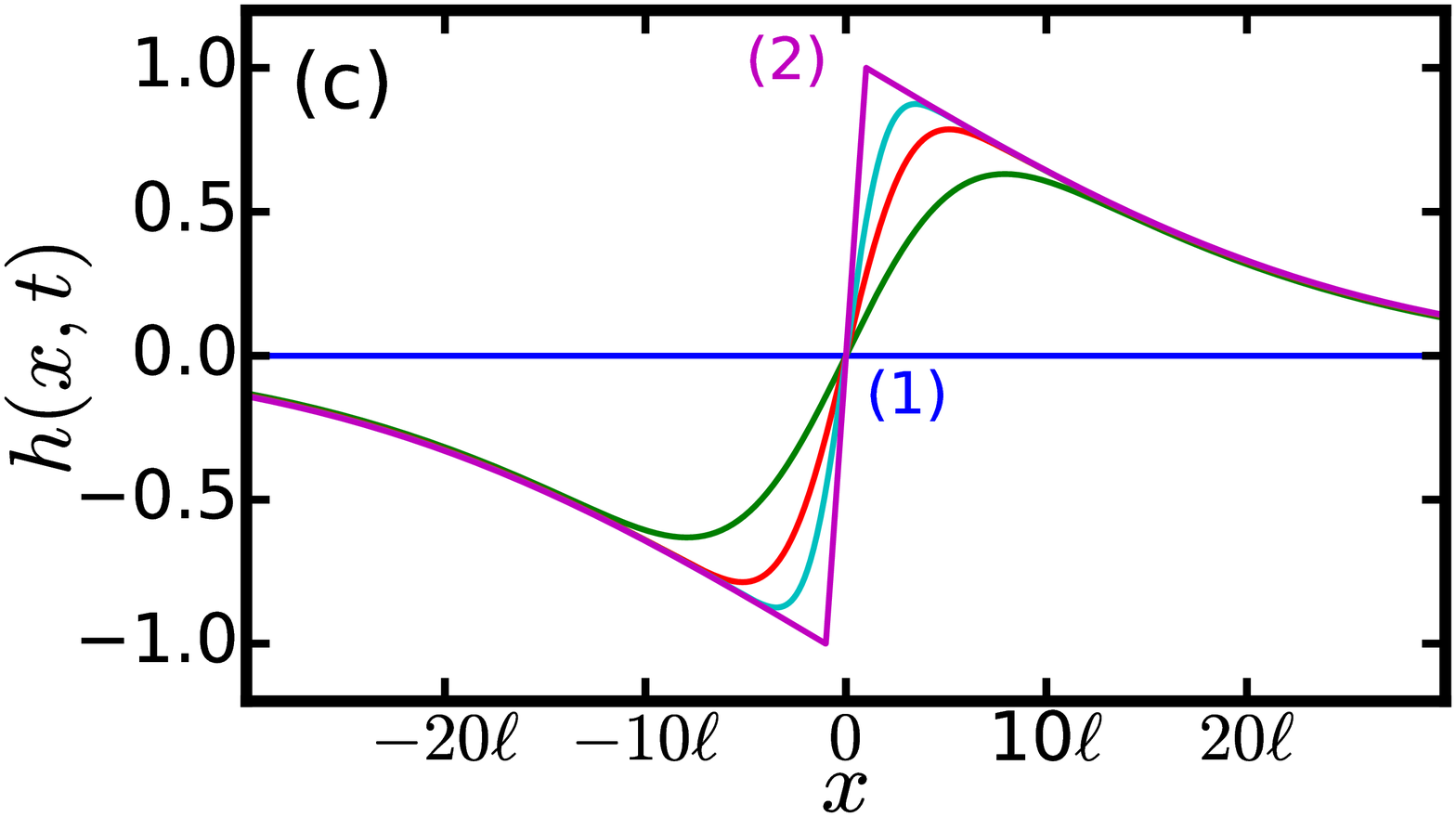}
\caption{The optimal path conditioned on reaching a rescaled height difference $\Delta=2$ at $t=T$ for the nonconserved EW equation (\ref{eq:EW_nc}).  The interface height $h(x,t)$ is rescaled by $D_0^{1/2}\nu^{-1/4}T^{1/4}$, $x$ is rescaled by $\sqrt{\nu T}$. $\ell = 10$ (a), $\ell = 1$ (b) and $\ell = 0.1$ (c). The initial ($t=0$) and final ($t=T$) profiles are marked by (1) and (2), respectively. On panels (a) and (b) the height is plotted at times $t=0$, $T/4$, $T/2$, $3T/4$, and $T$. On panel (c) the times are $t=0$, $0.9T$, $0.97T$, $0.99T$, and $T$.}
\label{fig:optimal_history_conserved}
\end{figure}

The optimal history $h\left(x,t\right)$ of the non-conserved one-dimensional EW equation (\ref{eq:EW_nc}), conditioned on a given height difference $\Delta$, is obtained by integrating Eq.~(\ref{eq:sol_q}) with respect to $x$ with the boundary conditions $h\left(x\to\pm\infty,t\right)\to0$.
It is given by:
\begin{eqnarray}
\label{eq:hsol_nonconserving}
h\left(x,t\right)\!\!&=&\!\!-\frac{\lambda}{4}\!\!\sum_{j_{1},j_{2}=\pm1}\!\!\!\!j_{1}j_{2}\!\left\{ \!\sqrt{\frac{4\left(1+j_{2}t\right)}{\pi}}\exp\!\left[\!-\frac{\left(x+j_{1}\ell\right)^{2}}{4\left(1+j_{2}t\right)}\right]\right. \nonumber\\
&+& \left.\left(x+j_{1}\ell\right)\text{erf}\left[\frac{x+j_{1}\ell}{\sqrt{4\left(1+j_{2}t\right)}}\right]\right\},
\end{eqnarray}
where $t$ and $x$ are rescaled as in Eq.~(\ref{eq:langevin_dimensionless}), but the interface height $h$ is rescaled by $\mu = D_0^{1/2}\nu^{-1/4}T^{1/4}$.
The value of $\lambda$ is found from Eq.~(\ref{sol_lambda}) with $\bar{h}$ replaced by $\Delta\sqrt{\nu T}/\left(2\mu L\right)$. Figure \ref{fig:optimal_history_conserved} depicts the optimal interface height histories $h(x,t)$ for the non-conserved EW equation, conditioned on a given height difference $\Delta$.

In the short-time limit, $\ell \gg 1$, the optimal interface dynamics~(\ref{eq:hsol_nonconserving}) are localized at the boundaries  $x=\pm\ell$. Around each of the boundaries, the optimal profile is well approximated by the solution to the \emph{one-point} problem: when we condition the process on reaching height $\pm\Delta /2$, respectively, at $t=T$. This can be seen by comparing the optimal path (\ref{eq:hsol_nonconserving}) around $x=\pm\ell$ with the optimal path of the one-point problem, see Eq.~(33) in Ref. \citep{MKV_PRL2016}. Taking the limit $L\to\infty$ in Eq.~(\ref{eq:Var_Delta_nonconserving}) we obtain the action in the physical units:
\begin{equation}
S=\frac{\Delta^{2}}{2\text{Var}\left(\Delta\right)}\simeq\frac{\Delta^{2}}{2D_{0}}\sqrt{\frac{\pi\nu}{2T}}.
\end{equation}
It is equal to twice the action evaluated on the solution to the one-point problem conditioned on reaching height $\Delta/2$, see Eq.~(39) in Ref. \cite{MKV_PRL2016}.

In the long-time limit, $\ell \ll 1$, the final optimal profile can be approximated, close to the origin, by
\begin{equation}
\label{eq:optimal_final_profile_nonconserving}
h\left(x,T\right)\simeq\begin{cases}
+\frac{\Delta}{2}, & x>L,\\
\frac{\Delta x}{2L}, & \left|x\right|\le L,\\
-\frac{\Delta}{2}, & x<-L,
\end{cases}\quad\left|x\right|,L\ll\sqrt{\nu T}
\end{equation}
(in the physical units).
As one can easily check, this profile minimizes the free energy
\begin{equation}
\label{eq:free_energy_nonconserved}
F_{\text{EW}}\left[h\right]=\frac{\nu}{D_{0}}\int dx\left(\partial_{x}h\right)^{2}
\end{equation}
of the nonconserved EW equation, as to be expected in thermal equilibrium. In its turn, the action in this limit,
\begin{equation}\label{equilactionEW}
S=\frac{\nu\Delta^2}{2 D_0L},
\end{equation}
coincides with the free energy (\ref{eq:free_energy_nonconserved})
evaluated on the optimal profile~(\ref{eq:optimal_final_profile_nonconserving}). These long-time results also hold for the KPZ equation in one dimension, since the free energy~(\ref{eq:free_energy_nonconserved}) yields the stationary probability distribution of height profiles in the KPZ equation as well \citep{HHZ,Barabasi}.

\section{KPZ Equation}

Until now we have been dealing with linear interface growth models described by Eqs.~(\ref{eq:generalized_EW}) and (\ref{eq:noise}). However, in the absence of a small-scale cutoff, the ill-posedness of the finite-time one-point height distribution also appears in nonlinear growth models. As an important example, let us consider the KPZ equation \cite{KPZ}
\begin{equation}
\label{eq:2DKPZ}
\partial_{t}h=\nu \nabla^{2} h+\frac{\lambda}{2}\left(\nabla h\right)^2 +\sqrt{D}\,\eta\left(\vect{x},t\right),
\end{equation}
which generalizes the nonconserved EW equation  ($m=1$, $\alpha=0$) by accounting for an important nonlinearity which breaks the up-down symmetry. In view of the forthcoming results let us introduce finite spatial correlations of the Gaussian noise
by replacing Eq.~(\ref{eq:noise}) (with $\alpha=0$) with the following one:
\begin{equation}\label{corrnoise}
\langle\eta(\mathbf{x}_{1},t_{1})\eta(\mathbf{x}_{2},t_{2})\rangle = C(|\mathbf{x}_{1}-\mathbf{x}_{2}|)\delta(t_{1}-t_{2}).
\end{equation}
We will represent the spatial correlator as $C(r) = \delta^{-d} c(r/\delta)$, where the volume integral of $C(r)$
is equal to $1$.  Sending the correlation length $\delta$ to zero, one restores the limit of the delta-correlated noise. Let us consider $d=2$. As previously, we start from a flat interface at $t=0$ and study the probability distribution to observe the interface height $H$ at the origin at time $t=T$.  The rescaling transformation
\begin{equation}\label{rescalingKPZ}
t/T\to t, \quad x/\sqrt{\nu T} \to x, \quad \lambda h/\nu\to h
\end{equation}
brings Eq.~(\ref{eq:2DKPZ}) to the following form:
\begin{equation}\label{KPZrescaledd}
\partial_{t}h=\nabla^2 h+\frac{1}{2}\left(\nabla h\right)^2+\sqrt{\epsilon} \,\eta(\mathbf{x},t),
\end{equation}
where $\epsilon=D\lambda^2/\nu^3$,
and we have assumed, without loss of generality, that $\lambda>0$. The rescaled correlation
length is $\delta/\sqrt{\nu T}$.  The (normalized to unity) one-point height distribution at the origin at time $T$ can depend only on
three dimensionless parameters:
$\tilde{H}=\lambda H /\nu$, $\epsilon$ and $\nu T/\delta^2$:
\begin{equation}\label{exactscaling}
{\mathcal P}(H,T) = \frac{\lambda}{\nu}\,f_{\epsilon}\left(\frac{\lambda H}{\nu},\frac{\nu T}{\delta^2}\right).
\end{equation}
Importantly, and exclusively for $d=2$,
the time-dependence in the right-hand-side of the exact Eq.~(\ref{exactscaling}) appears only through the combination $\nu T/\delta^2$
which includes the correlation length of the noise \cite{units}.

In the weak-coupling regime, $\epsilon\ll 1$, the body of the height distribution is Gaussian,
as described by the EW equation, until exponentially long times \cite{Natterman1992}.
As explained in Sec. II, the height-distribution variance diverges here as $\delta \to 0$.
The height-distribution tails are also ill-defined at $\delta \to 0$ at $d\geq 2$, as follows from the
WNT of the KPZ equation \cite{KK2008}.

What happens in the strong-coupling regime, $\epsilon\gg 1$? In the absence of analytic results for ${\mathcal P}(H,T)$, Halpin-Healy \cite{Halpin2012,Halpin2013} performed extensive numerical simulations of this regime for $d=2$. He simulated both the KPZ equation itself, and several discrete models, believed to
belong to the KPZ universality class. He did it for different initial conditions (including the flat one), which do not introduce a macroscopic length scale.
He clearly observed, for all these models (and at long but not too long times, when the system size is still irrelevant), a universal self-similar distribution
of the typical fluctuations of the interface height.  According to Refs. \cite{Halpin2012,Halpin2013}, this self-similar distribution can be represented as
\begin{equation}\label{HHscaling}
{\mathcal P} (H,T)= \frac{1}{c_1\, T^{\beta}}\, g \left(\frac{H-c_2 T}{c_1\,T^{\beta}} \right),
\end{equation}
where $\beta \simeq 0.240$ in agreement with earlier simulations, and the constant coefficients $c_1$ and $c_2$ are model-dependent. As observed
in Refs.~\cite{Halpin2012,Halpin2013}, for a proper choice of $c_1$ and $c_2$, the function $g$ is universal. The self-similar behavior of the height distribution (\ref{HHscaling})
is of the same type as the one rigorously established (at long times, for typical fluctuations, and for several types of initial
conditions) at $d=1$ \citep{Corwin,Quastel2015,HHT,Spohn2016}.

Equations~(\ref{exactscaling}) and (\ref{HHscaling}) are compatible only if, for typical fluctuations,
\begin{equation}\label{jointscaling}
{\mathcal P}(H,T) = \frac{\lambda}{\nu\, C_1(\epsilon)}\, \left(\frac{\delta^2}{\nu T}\right)^\beta\,  F \left[\frac{\frac{\lambda H}{\nu}-C_2(\epsilon) \frac{\nu T}{\delta^2}}{C_1(\epsilon)\, \left(\frac{\nu T}{\delta^2}\right)^{\beta}}\right],
\end{equation}
where $F(w)$ is a universal function of its single argument, and $C_1$ and $C_2$ are universal functions of $\epsilon$
up to numerical coefficients that can be model-dependent.

According to Eq.~(\ref{jointscaling}), the standard deviation of height from its mean behaves as \cite{largeepsilon}

\begin{equation}\label{varKPZ}
\sqrt{\left\langle h^{2}\left(0,T\right)\right\rangle -\left\langle h\left(0,T\right)\right\rangle ^{2}} \sim \frac{\nu C_1(\epsilon)}{\lambda} \left(\frac{\nu T}{\delta^2}\right)^{\beta}.
\end{equation}
This quantity clearly diverges, and the one-point height distribution becomes ill-defined, in the limit of $\delta \to 0$, that is for white spatial noise. As one can see from Eq.~(\ref{varKPZ}), the KPZ nonlinearity \emph{amplifies} the UV catastrophe of the EW equation at $d=2$: the divergence becomes a power-law (rather
than logarithmic) in $\delta$. The systematic interface velocity \cite{largeepsilon},
\begin{equation}\label{VKPZ}
\mathcal{V}=\frac{\nu^2 C_2(\epsilon)}{\lambda \delta^2},
\end{equation}
which results from the rectification of the noise by the nonlinearity, also diverges as $\delta \to 0$, but a similar divergence  occurs already at $d=1$, see \textit{e.g.} Ref. \cite{Hairer}.

In numerical simulations there is always a small-scale cutoff, such as the grid size in numerical integration schemes, the lattice constant in discrete models, etc. Still, one should remember that the \emph{amplitudes} of
the scaling relations, stemming from Eq.~(\ref{jointscaling}), such as Eq.~(\ref{varKPZ}), are non-universal: they are determined by a system-dependent small-scale cutoff. The far tails of the distribution, not necessarily described by Eq.~(\ref{HHscaling}), also depend on the small-scale cutoff. It would be interesting to explore whether the local-average-height statistics provides a viable alternative.

\section{Summary and Discussion}

For every interface growth model, described by Eqs.~(\ref{eq:generalized_EW}) and~(\ref{eq:noise}), there is a critical dimension $d_c$ (\ref{eq:critical_dimension}), at or above which the finite-time one-point height distribution is ill-defined
because of a UV catastrophe.
Here we introduced a macroscopic regularization of this catastrophe, by shifting the
attention to
the local average height (\ref{eq:hbar_general_dim}). For Eqs.~(\ref{eq:generalized_EW}) and~(\ref{eq:noise}) the distribution of this quantity is well-defined in any dimension
without need for a regularization of the model at small scales.  We calculated the variance of this (Gaussian) distribution for all models
described by Eq.~(\ref{eq:generalized_EW}) and for all dimensions, see Eq.~(\ref{eq:variance_exact}).  In addition, we formulated the weak-noise theory (WNT) which allows one to determine the optimal path of the system: the most probable history of the interface conditioned on a given value of the local average height $\bar{h}$ at a specified time. We performed explicit calculations for the simple case of the conserved EW equation in 1+1 dimensions (\ref{eq:EW}).  We then used these results to study the distribution of the height difference in the  nonconserved EW equation, and to determine the optimal path given such a height difference.

The ill-posedness of the finite-time one-point height distribution at $d\geq d_c$
also appears in nonlinear interface models
without a small-scale cutoff, for example, for the KPZ equation in $2+1$ dimensions.
As we argue, the amplitudes of scaling relations, uncovered in  numerical simulations at $d\geq d_c$, depend on
an (explicit or implicit) small-scale cutoff. Moreover, the nonlinearity significantly changes the cutoff dependence of the amplitudes compared with the non-conserved
EW-equation in $2+1$ dimensions. It would be interesting to explore whether, and under which conditions, the local average height provides a viable regularization alternative to the small-scale cutoff in nonlinear models.

When the noise is \emph{typically} weak, the
statistics of the local average height can be probed using
the weak-noise theory (WNT). For nonlinear  models the WNT equations are much harder to solve than the simple linear equations that we analyzed here. Still, useful analytic asymptotics for the optimal path and the action can be found in different limits
and for different initial conditions, as has been shown for $d=1$ in Refs.
\citep{KK2007,KK2008,MKV_PRL2016,KMS2016,Janas2016,KK2009}.  Also, an efficient numerical algorithm for solving
the WNT equations
is available \citep{CS,Grafke}. It would be interesting to use the WNT for determining the far tails of the
distribution of the local average height in the KPZ equation at $d\ge 2$ and small $\epsilon$.

\section*{Acknowledgments}

We are very grateful to Joachim Krug for valuable advice and to Arkady Vilenkin for discussions.  N.S. and B.M.
acknowledge financial support from the Israel Science Foundation (grant No. 807/16)
and from the United States-Israel Binational Science Foundation (BSF) (grant No. 2012145).
B.M. also acknowledges support from the University of
Cologne through the Center of Excellence ``Quantum Matter and Materials."

\appendix
\section{One-point height distribution}
\label{appendix:single_point}

Consider the Fourier transform of the height-height correlation function (\ref{corr}):
 \begin{eqnarray}
\label{eq:F_k1_k2_def}
F\left(\vect{k}_{1},\vect{k}_{2},t\right)\!\!&=&\frac{1}{\left(2\pi\right)^{d}}\int d\vect{x}_{1} \, e^{i\vect{k}_{1}\cdot\vect{x}_{1}}\int d\vect{x}_{2} \, e^{i\vect{k}_{2}\cdot\vect{x}_{2}}\nonumber\\
 &\times& C\left(\vect{x}_{1},\vect{x}_{2},t\right)=\nonumber\\
 &=&\!\! \frac{1}{\left(2\pi\right)^{d}}\left\langle \int d\vect{x}_{1} \, e^{i\vect{k}_{1}\cdot\vect{x}_{1}}\int d\vect{x}_{2} \, e^{i\vect{k}_{2}\cdot\vect{x}_{2}}\right. \nonumber\\
 &\times& \! h\left(\vect{x}_{1},t\right)h\left(\vect{x}_{2},t\right)\biggr\rangle = \left\langle h\left(\vect{k}_{1},t\right)h\left(\vect{k}_{2},t\right)\right\rangle .\nonumber\\
\end{eqnarray}
It is easy to show that the Fourier transform of the noise (\ref{eq:noise_divergence}) satisfies the equation
\begin{eqnarray}
\left\langle \eta\left(\vect{k}_{1},t_{1}\right)\eta\left(\vect{k}_{2},t_{2}\right)\right\rangle \! &=& \! \left\langle i\vect{k}_{1}\cdot\vect{\xi}\left(\vect{k}_{1},t_{1}\right) \, i\vect{k}_{2}\cdot\vect{\xi}\left(\vect{k}_{2},t_{2}\right)\right\rangle \nonumber\\
&=& \! -k_{1}^{2\alpha}\delta\left(t_{1}-t_{2}\right)\delta\left(\vect{k}_{1}+\vect{k}_{2}\right). \nonumber\\
\end{eqnarray}
Using Eq.~(\ref{eq:generalized_EW}), we see that the Fourier transform of the height function,
\begin{equation}
h\left(\vect{k},t\right)=\frac{1}{\left(2\pi\right)^{d/2}}\int d\vect{x} \, e^{i\vect{k}\cdot\vect{x}} \,h\left(\vect{x},t\right),
 \end{equation}
obeys the equation
\begin{equation}
\label{eq:h_k_dynamics}
\partial_{t}h\left(\vect{k},t\right)=-\left(k^{2}\nu\right)^{m}h\left(\vect{k},t\right)+\sqrt{D} \, \eta\left(\vect{k},t\right).
 \end{equation}
The solution to this equation is
\begin{equation}
\label{eq:h_k_sol}
h\left(\vect{k},t\right)=\sqrt{D}\int_{0}^{t} dt' \, e^{-\left(k^{2}\nu\right)^{m}\left(t-t'\right)}\eta\left(\vect{k},t'\right).
\end{equation}
Then Eqs.~(\ref{eq:F_k1_k2_def}) and (\ref{eq:h_k_sol}) yield
\begin{equation}
\label{eq:F_k1_k2}
\!\!\! F \! \left(\vect{k}_{1},\vect{k}_{2},t\right)= \! D\,\delta\left(\vect{k}_{1}+\vect{k}_{2}\right)\!\frac{k_{1}^{2\alpha-2m}}{2\nu^{m}} \! \left[1-e^{-2\left(k_{1}^{2}\nu\right)^{m}t}\right] \! .
\end{equation}
The inverse Fourier transform of Eq.~(\ref{eq:F_k1_k2}) gives the correlation function (\ref{eq:correlation_function}) in physical space.

\section{The local average height}
\label{appendix:local_average_height}
The variance of the local average height (\ref{eq:hbar_general_dim}) (the averaging is performed over a $d$-dimensional hypercube) can be expressed through the Fourier transform of the interface height (\ref{eq:F_k1_k2_def}):
\begin{eqnarray}
\label{eq:appendix_var_hbar1}
\!\!\!\text{Var}\left[\bar{h}\left(t\right)\right] \!\!\!  &=& \!\!\int_{\left[-L,L\right]^{d}}\!\!\!\!\!d\vect{x}_{1}\int_{\left[-L,L\right]^{d}}\!\!\!\!\!d\vect{x}_{2}\,h\left(\vect{x}_{1},t\right)h\left(\vect{x}_{2},t\right) \nonumber\\
&=&  \!\!\frac{1}{\left(2L\right)^{2d}\left(2\pi\right)^{d}}\int d\vect{k}_{1}\int d\vect{k}_{2} \, F\left(\vect{k}_{1},\vect{k}_{2},t\right) \nonumber\\
&\times&   \!\!\!\int_{\left[-L,L\right]^{d}}\!\!\!\!\!d\vect{x}_{1} \, e^{-i\vect{k}_{1}\cdot\vect{x}_{1}}\!\!\int_{\left[-L,L\right]^{d}}\!\!\!\!\!d\vect{x}_{2} \, e^{-i\vect{k}_{2}\cdot\vect{x}_{2}}.
\end{eqnarray}
Evaluating the spatial integrals and using Eq.~(\ref{eq:F_k1_k2}), one arrives at Eq.~(\ref{eq:variance_exact}).

\section{Derivation of the WNT equations}
\label{appendix:WNT}

\subsection{Conserved Noise}

When the noise is conserved ($\alpha=1$), Eq.~(\ref{eq:generalized_EW}) can be written as
\begin{equation}
\label{eq:appendix_generalized_EW_conserved}
\partial_{t}h=-\left(-\nu\nabla^{2}\right)^{m}h+\sqrt{D}\,\nabla\cdot\vect{\xi}\left(\vect{x},t\right).
\end{equation}
We are interested in the probability of transition from the flat initial state, $h(\vect{x},t=0)=0$, to a state $h(\vect{x},t=T)$ with local average height $\bar{h}$.
Following Ref. \citep{KMSvoid}, we introduce a potential $\vect{u}\left(\vect{x},t\right)$ so that $h=\nabla\cdot\vect{u}$.
 Now
Eq.~(\ref{eq:appendix_generalized_EW_conserved}) becomes $\partial_{t}\vect{u}+\left(-\nu\nabla^{2}\right)^{m}\vect{u}=\sqrt{D}\,\vect{\xi}$.
The Gaussian action can be written as
\begin{eqnarray}
\label{eq:appendix_action}
S&=&\int_{0}^{T}\!\!\!dt\int\!d\vect{x}\frac{D\xi^{2}\left(\vect{x},t\right)}{2}\nonumber\\
&=& \int_{0}^{T}\!\!\!dt\int\!d\vect{x}\frac{\left[\partial_{t}\vect{u}+\left(-\nu\nabla^{2}\right)^{m}\vect{u}\right]^{2}}{2}.
\end{eqnarray}
In the spirit of weak-noise theory, we must minimize this action with respect to the trajectory $\vect{u}\left(\vect{x},t\right)$. Introduce the ``momentum density" field $p(\vect{x},t)$ and a solenoidal vector field $\vect{\omega}$, $\nabla\cdot\vect{\omega} = 0$, so that \citep{KMSvoid}
\begin{equation}
\label{eq:appendix_pdef}
\partial_{t}\vect{u}+\left(-\nu\nabla^{2}\right)^{m}\vect{u}=-\nabla p+\vect{\omega}.
\end{equation}
Plugging Eq.~(\ref{eq:appendix_pdef}) into Eq.~(\ref{eq:appendix_action}) and integrating by parts we obtain
\begin{equation}
\label{eq:appendix_action_with_omega}
\!\! S=\! \int_{0}^{T}\!\!\!dt \! \int \! d\vect{x}\left[\frac{\left(\nabla p\right)^{2}}{2}+\frac{\vect{\omega}^{2}}{2}\right]- \int_{0}^{T}\!\!\!dt \! \int \! d\vect{x}\,\nabla\cdot\left(p\vect{\omega}\right).
\end{equation}
By virtue of the Gauss theorem, together with the boundary conditions at $\left|\vect{x}\right|\to\infty$, the last term in Eq.~(\ref{eq:appendix_action_with_omega}) vanishes. Therefore, the action is minimum for $\vect{\omega} = 0$. $p$ is then given by the solution to the Poisson equation
\begin{equation}
\label{eq:appendix_poisson_equation_p}
\nabla^{2}p=-\nabla\cdot\left[\partial_{t}\vect{u}+\left(-\nu\nabla^{2}\right)^{m}\vect{u}\right]
\end{equation}
with the boundary condition $p\left(\left|\vect{x}\right|\to\infty,t\right)\to0$, and the action is given by
\begin{equation}
S=\int_{0}^{T}dt\int d\vect{x}\,\frac{\left(\nabla p\right)^{2}}{2}.
\end{equation}
The variation of this action is
\begin{equation}\label{variation}
\delta S\left[\vect{u}\right]=-\int_{0}^{T} \!\!\! dt\int \! d\vect{x}\,\nabla p \, \cdot\left[\partial_{t}\delta\vect{u}+\left(-\nu\nabla^{2}\right)^{m}\delta\vect{u}\right].
\end{equation}
After several integrations by parts we obtain
\begin{eqnarray}
\label{eq:appendix_variation_with_BC}
\delta S & = &\int_{0}^{T} \! dt\int d\vect{x}\left[-\partial_{t}p+\left(-\nu\nabla^{2}\right)^{m}p\right]\left(\nabla\cdot\delta\vect{u}\right) \nonumber\\
&+& \int d\vect{x}\left[p\,\nabla\cdot\delta\vect{u}\right]_{t=0}^{t=T}.
\end{eqnarray}
The bulk term, which comes from the demand that $\delta S$ vanish for arbitrary $\delta\vect{u}$, yields the first WNT equation:
\begin{equation}\label{eq:peq_appendix}
\partial_{t}p=\left(-\nu\nabla^{2}\right)^{m}p.
\end{equation}
Equation~(\ref{eq:appendix_poisson_equation_p}) yields the second WNT equation:
\begin{equation}\label{eq:heq_appendix}
\partial_{t}h=-\left(-\nu\nabla^{2}\right)^{m}h-\nabla^{2}p.
\end{equation}
Eqs.~(\ref{eq:peq_appendix}) and (\ref{eq:heq_appendix}) are to be solved together with the initial condition $h\left(\vect{x},t=0\right)=0$. The condition on the local average height at $t=T$ introduces an integral constraint (\ref{eq:hbar_general_dim}), which calls for a Lagrange multiplier \citep{DG2009}. As a result, an additional term,
\begin{equation}
\Lambda\int_{\Omega}\!\!\!d\vect{x}\,h\left(\vect{x},T\right)=\Lambda\int_{\Omega}\!\!\!d\vect{x}\,\nabla\cdot\vect{u}\left(\vect{x},T\right)
\end{equation}
should be added to the action functional. Together with the terms in the right-hand-side of Eq.~(\ref{eq:appendix_variation_with_BC}), this term yields the boundary condition (\ref{eq:initial_condition_p_general}) on the momentum density at $t=T$.

\subsection{Non-Conserved Noise}

The case of non-conserved noise is more straightforward. Consider Eq.~(\ref{eq:generalized_EW}) with $\alpha=0$,
\begin{equation}
\partial_{t}h=-\left(-\nu\nabla^{2}\right)^{m}h+\sqrt{D}\,\xi\left(\vect{x},t\right),
\end{equation}
and define the momentum density $p\equiv  \partial_{t}h + \left(-\nu\nabla^{2}\right)^{m}h$. Then the Gaussian action is:
\begin{equation}
 S=\int_{0}^{T}dt\int d\vect{x} \, \frac{p^{2}}{2}.
\end{equation}
The variation of the action is, after integrating by parts,
\begin{eqnarray}
\delta S&=&\left[\int\!d\vect{x}\,p\,\delta h\right]_{t=0}^{t=T}-\int_{0}^{T}\!dt\int\!d\vect{x}\,\partial_{t}p\,\delta h \nonumber\\
&+&\int_{0}^{T}\!dt\int\!d\vect{x}\left(-\nu\nabla^{2}\right)^{m}p \, \delta h.
\end{eqnarray}
The bulk term yields the WNT equation
\begin{equation}
 \partial_{t}p=\left(-\nu\nabla^{2}\right)^{m}p.
\end{equation}
The second WNT equation,
 \begin{equation}
\partial_{t}h=-\left(-\nu\nabla^{2}\right)^{m}h+p,
\end{equation}
follows immediately from the definition of $p$.
As in the conserved case, the boundary condition on the momentum density at $t=T$, due to the constraint on the local average height,
is given by Eq.~(\ref{eq:initial_condition_p_general}).\\

\section{Evaluation of the Action for the Non-Conserved EW Equation}
\label{appendix:erf_integral}

Plugging Eq.~(\ref{eq:sol_p}) into Eq.~(\ref{eq:action_space_integral}) yields:
\begin{equation}
\label{eq:S_appendix1}
s=\frac{\lambda^{2}\ell}{16}\left[8-J\left(\ell\right)\right],
\end{equation}
where
\begin{equation}
\label{eq:J_def}
J\left(\ell\right)=\int_{-\infty}^{\infty} \! dz\left[\text{erf}\left(\ell\frac{z+1}{2}\right)-\text{erf}\left(\ell\frac{z-1}{2}\right)\right]^{2}.
\end{equation}
To evaluate this integral, we first evaluate its derivative with respect to $\ell$:
\begin{eqnarray}
\label{eq:dJ_dl}
\frac{d J}{d\ell} &=& \frac{2}{\sqrt{\pi}}\int_{-\infty}^{\infty} \! dz\left[\text{erf}\left(\ell\frac{z+1}{2}\right)-\text{erf}\left(\ell\frac{z-1}{2}\right)\right] \nonumber\\
&\times&\left[\left(z+1\right)e^{-\frac{1}{4}\ell^{2}\left(z+1\right)^{2}}-\left(z-1\right)e^{-\frac{1}{4}\ell^{2}\left(z-1\right)^{2}}\right]\nonumber\\
&=& 8\sqrt{\frac{2}{\pi}}\left(\frac{1-e^{-\ell^{2}/2}}{\ell^{2}}\right).
\end{eqnarray}
Now we integrate Eq.~(\ref{eq:dJ_dl}) over $\ell$, taking into account the condition $J\left(\ell=0\right)=0$ which follows from Eq.~(\ref{eq:J_def}).
We obtain
\begin{equation}
\label{eq:J_sol}
J\left(\ell\right)=8\left[\sqrt{\frac{2}{\pi}} \, \frac{e^{-\ell^{2}/2}-1}{\ell}+\text{erf}\left(\frac{\ell}{\sqrt{2}}\right)\right].
\end{equation}
Plugging Eqs.~(\ref{sol_lambda}) and (\ref{eq:J_sol}) into Eq.~(\ref{eq:S_appendix1}) we arrive at Eq.~(\ref{eq:S_sol}).


\begin{thebibliography}{99}
\bibitem{units} Another way of saying it is that, for $d=2$ and $\delta=0$, Eq.~(\ref{eq:2DKPZ}) does not have an intrinsic length or time scale. To clearly see the latter point one should attribute independent units to the interface height $h$.
\bibitem{largeepsilon}  At $\epsilon \gg 1$ the nonlinear term
    in the KPZ equation dominates over the diffusion term, the latter only being important in the regions of attempted
    cusp formation. Therefore,  it is reasonable to assume that the expressions in the right-hand-sides of Eqs.~(\ref{jointscaling})-(\ref{VKPZ}) become independent of $\nu$, and we obtain $C_1(\epsilon)=\bar{C}_1\, \epsilon^{(1+\beta)/3}$, and $C_2(\epsilon)=\bar{C}_2\, \epsilon^{2/3}$, where $\bar{C}_1\sim 1$ and $\bar{C}_2\sim 1$ are model-dependent constants. This argument applies only at $d=2$, where there are no intrinsic length and time scales in the problem except those determined by the small-scale cutoff $\delta$.
\end{thebibliography}
\end{document}